# STEERED-MIXTURE-OF-EXPERTS REGRESSION FOR IMAGE DENOISING WITH MULTI-MODEL INFERENCE

*Aytac Özkan, Yi-Hsin Li and Thomas Sikora*

Technische Universität Berlin, Germany

## ABSTRACT

In this paper we introduce a novel block-based regression strategy for image denoising based on edge-aware Steered-Mixture-of-Experts (SMoE) models. SMoEs provide very sparse image representations – able to model sharp edges as well as smooth transitions in images efficiently with few parameters. A multi-model inference strategy is developed that improves significantly the denoising capacity of single SMoE models. We show that the important edge reconstruction properties of SMoEs are well preserved, even when many models are fused under severe noise. We investigate model-inference from local neighborhood blocks as well as from distant blocks using block-matching as in BM3D. Our initial results indicate that SMoE multi-model regression can provide promising results compared to state-of-the-art BM3D with excellent edge quality.

**Index Terms—** Image denoising, Steered-Mixture-of-Experts, multi-model inference, regression model

## I. INTRODUCTION

Image denoising is an important task in essentially all image application areas and a very active field of research. The last 30 years witnessed tremendous progress with classical approaches. Prominent strategies include spatial-domain algorithms based on total variation regularization (TV) [1][2][3], non-local means (NLM) [4][5] as well as transform domain techniques. Here the block-matching 3D transform (BM3D) approach [6][7] with extensions to volumetric data [8] and light fields [9] stands out with its excellent performance. The reader is referred to [10][11] for an overview on the topic. Most recently, "learnt" denoising approaches based on deep neural networks have demonstrated impressive progress [12]. An excellent overview of state-of-the art methods is provided with the NITRE 2020 challenge [13], which is based on a dataset in [14]. The classical BM3D approach still provides competitive performance in this context.

In this paper we introduce "Steered Mixture of Experts" (SMoE) models for denoising. SMoEs are edge-reconstructing gating networks and have previously been explored for compression of images, video and lightfields [15-23]. One intriguing property of SMoEs is sparsity: the ability to reproduce sharp edges (discontinuities) as well as smooth transitions in images with few parameters. Sparsity and edge-awareness are important prerequisites for compression, but also of vital importance when used to recover structures like edges from noise without overfitting. We use a regression framework to fit SMoE models to pixels in small image blocks. To boost performance, we devise a multi-model inference approach to fuse outcomes of $H$ SMoE models from either local or global block neighborhoods. Our motivation to use small blocks and to fuse multiple blocks for denoising is twofold:

- To be compatible to block sizes in 3D/4D groups of the BM3D/BM4D algorithms [6][7]. Our hypothesis is, that SMoEs on BM3D blocks with significant edge structure can improve over DCT denoising in BM3D.
- Fitting SMoE models is computationally demanding. Recent work in [24] indicates that SMoE parameters are efficiently derived on small blocks using ultra-fast Autoencoder "unfolding". This will avoid in the future cumbersome iterative optimization to meet real-time demands.

Using regression strategies for denoising has a long history. Recent work can be found in [25-29]. Most strategies are generally not competitive to state-of-the-art BM3D. Related work includes denoising with Kernel Support Vector Machines (SVM) [26][27] and Kernel Regression (KR) [28][29]. The underlying model of both SVM and KR is a single-pixel regressor with "weighted sum of kernels" which acts in a local neighborhood. The SMoE model in contrast is block-based and builds on a "weighted sum of soft-gated kernels". This is a novel model with edge-capability through gating with global pixel support.

## II. THE EDGE-AWARE SMoE MODEL

We model image pixel values at location $\underline{x}$ based on a Mixture-of-Experts with $L$ modes [21]:

$$y(\underline{x}) = \sum_{j=1}^{L} m_j(\underline{\mathrm{x}}) \cdot w_j(\underline{x}) \qquad (1)$$

$m_j(\underline{x})$ is an „expert" function responsible to explain the data in the region $R_j$ defined by "gating function" $w_j(\underline{x})$. We assume $\sum_{j=1}^{L} w_j(\underline{x}) = 1$. Thus, $w_j(\underline{x})$ defines the probability $P(R_j|\underline{X} = \underline{x}_k)$ that the expert $m_j(\underline{x}_k)$ is responsible for explaining the data at location $\underline{x}_k$. Notice, that $y(\underline{x})$ is a continuous function in $\underline{x}$. Once the edge-aware model function has been estimated from noisy data, it can be re-sampled to any resolution to readily provide i.e. edge-aware super-resolution representations as in [26]. In general, $\underline{x} \in R^n$ for n-dimensional imagery. $x \in R^1$ accounts for i.e. *1D* pixel scan lines, $\underline{x} \in R^2$ for *2D* gray-level images.

This work has received funding from the European Union's Horizon 2020 research and innovation programme under the Marie Skłodowska-Curie grant agreement No 956770, PLENOPTIMA.

For use in our regression framework, we convert the above general model in (1) into a parametric model [18][21]. Here we assume that the "experts" are constants $m_j(\underline{x})$=$m_j$ [20][22][24] and the gating functions $w_j(\underline{x})$ are derived via kernel functions $K_j(\underline{x})$ in the form

$$w_j(\underline{x}) = \frac{\pi_j \cdot K_j(\underline{x})}{\sum_{l=1}^{L} \pi_l \cdot K_l(\underline{x})} \quad \text{with} \quad \sum_{j=1}^{L} \pi_j = 1$$

Each kernel $K_j(\underline{x})$ is defined as "Steered Gaussian"

$$K_j(\underline{x}) = e^{-\left\{(\underline{x}-\underline{c}_j)^{\boldsymbol{T}} \cdot \underline{\Sigma}_j^{\boldsymbol{-1}} \cdot (\underline{x}-\underline{c}_j)\right\}}$$

with $\underline{c}_j^T$ the „center vector" (position of kernel). $\underline{\Sigma}_j \in R^{n \times n}$ is the kernel "steering" matrix, describing the bandwidth $\delta_{x_{il}}^2$ of the kernel in each dimension and $s_{il}, (i \neq l)$ the steering properties

$$\underline{\Sigma}_j = \begin{bmatrix} \delta_{x_{11}}^2 & s_{12} \cdots & s_{1n} \\ s_{21} & \ddots & \vdots \\ \vdots & \cdots & \delta_{x_{nn}}^2 \end{bmatrix}$$

The SMoE "steering" kernel regression model is thus parametric with parameters $\underline{\theta}$:

$$y(\underline{x}) = \sum_{j=1}^{L} \frac{m_j \cdot \pi_j \cdot e^{-\left\{(\underline{x}-\underline{c}_j)^{\boldsymbol{T}} \cdot \underline{\Sigma}_j^{\boldsymbol{-1}} \cdot (\underline{x}-\underline{c}_j)\right\}}}{\sum_{l=1}^{K} \pi_j \cdot e^{-\left\{(\underline{x}-\underline{c}_l)^{\boldsymbol{T}} \cdot \underline{\Sigma}_l^{\boldsymbol{-1}} \cdot (\underline{x}-\underline{c}_l)\right\}}} = f(\underline{x}, \underline{\theta}) \qquad (2)$$

This "steering" model degenerates into a "non-steering model if each full-parametric steering matrix $\underline{\Sigma}_j$ is converted into a diagonal matrix, or even into $\underline{\Sigma}_j = \delta_x^2 \cdot \underline{I}$, with $\underline{I}$ identity matrix. In the latter case the kernels are round, concentric kernels as used in [24].

For denoising an image block, the optimal SMoE-model parameters $\underline{\theta}_{opt}$ are first estimated from $M$ discrete noisy block pixels $D = \{y_r(\underline{x}_m), \underline{x}_m\}_{m=1}^{M}$. Subsequently the optimized model is used to estimate (predict) each pixel value $y(\underline{x})$ from the inference of the model $\hat{y}(\underline{x}) = f(\underline{x}, D, \hat{\underline{\theta}}_{opt})$. We emphasize, that the data $D$ doesn't have to be located on a regular grid and can be irregularly sampled from any arbitrarily shaped region in $R^n$.

A variety of optimization algorithms can be employed for parameter estimation, including Maximum-Expectation (ME) optimization [21] and Gradient Descent (GD) [18,19,22,24] procedures. Previous work [19] has shown that GD algorithms result in much improved edge reconstruction compared to ME approaches. In addition, with GD it is possible to optimize models with respect to very flexible objective criteria, including mean-squared-error and structural similarity index measure (SSIM) with and without side conditions [22]. SMoE Autoencoders enable ultra-fast SMoE parameter estimation [24].

## III. SMoE REGRESSION FOR DENOISING

Since the SMoE model has not been used as yet for denoising applications it is beneficial to illustrate the edge-aware model capabilities and advantages using basic examples. In Fig. 1 the denoising of a $M$ = *32x32* gray-level image-block using GD optimization is shown. A sparse model with only $L$=*10* steered kernels was employed for modeling the pixel values. The SMoE model is able to preserve the sharp edges as well as the smooth transitions in the image with excellent quality for perfect denoising. By estimating the steering properties of the kernels, the soft-gating functions $w_j(\underline{x})$ are derived, which results in the separation of the image into 10 soft-gated regions. Sharp transitions of gates into neighboring regions provide for the reconstruction of sharp edges, and soft-transitions of gates for smooth transitions (right side in (d)). In essence, during the optimization of parameters, kernels compete for "harvesting" the correlation of pixel values in their neighborhood for denoising. After optimization, all kernels collaborate for the reconstruction of $y(\underline{x})$.

Fig. 2 illustrates this on an artificial 1D pixel signal. A SMoE model with $L = 3$ kernels was used for modeling and also for denoising. True and estimated locations of the kernels are depicted in Fig. 2 as tuples $\{c_j, m_j\}$. With a block-size of $M$=*3000* samples (left) the model parameters as well as the signal itself are reconstructed from noise with excellent accuracy. Even though the signal is heavily contaminated with noise. From inspection of the gating functions it is apparent how the partitioning of the signal into sharp and smooth regions results. In practice, reconstruction from a small number of Nyquist-sampled signals is commonly required. Unfortunately, estimating the model parameters from such limited amounts of noisy samples causes model errors which cannot be neglected. The "small sample" problem is depicted in Fig. 2 (middle) based on $M$=*30* pixels. The scale of the noise signal $\varepsilon(x)$ is now closer to the scale of the original signal $y(x)$ and the model parameters and thus the signal itself is estimated with offsets. Notice, however, that the steep edge and smooth transitions are still recovered very well from noise and generally the shape of the signal is well preserved.

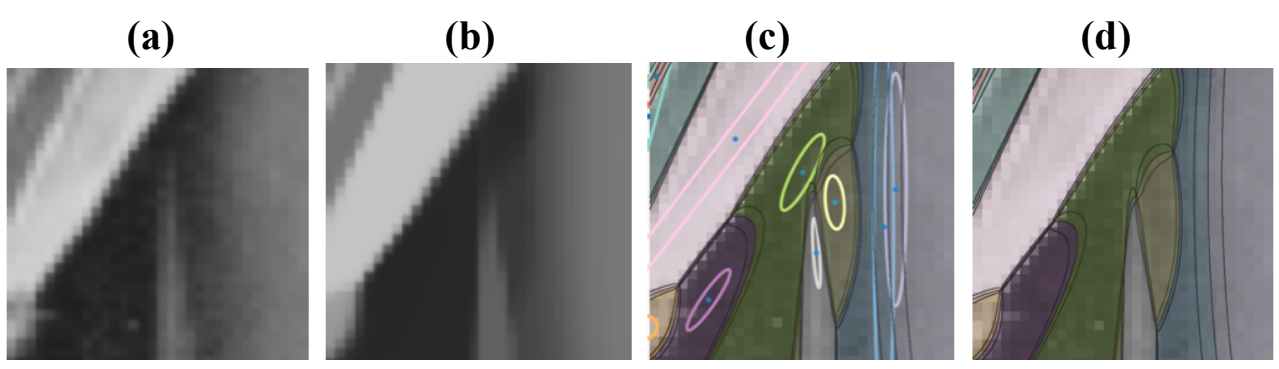

**Fig. 1.** SMoE denoising: (a) noisy patch (b) denoising result (c) Steered Gaussians (d) 2D gating functions $w_j(\underline{x})$ [24].

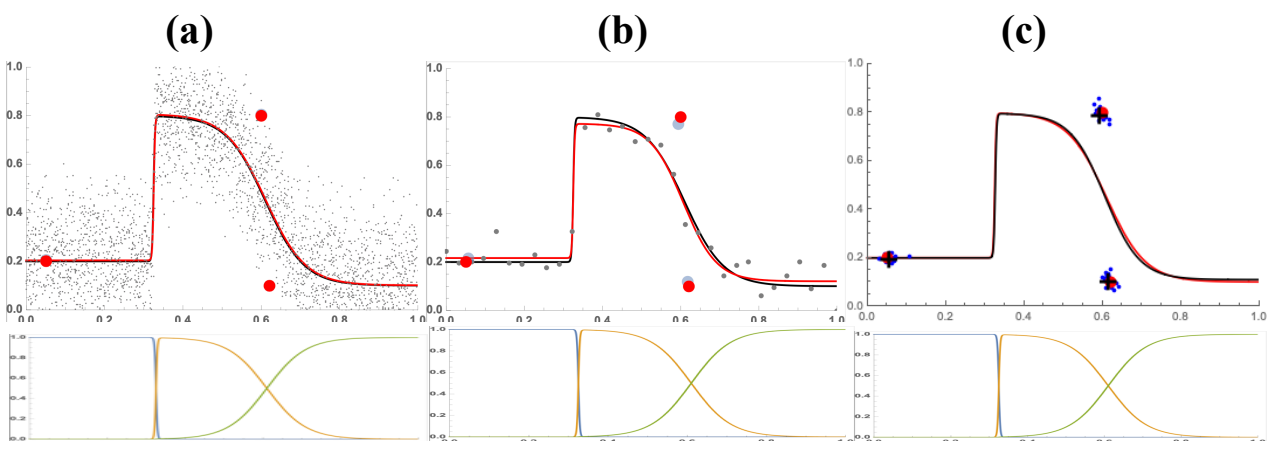

**Fig. 2.** Small sample problem: (a): Original and denoised 1D signal in a block of M=3000 samples with 3 kernels (red), and gating functions, $\delta_\varepsilon^2 = 0.15$. (b): Reconstructed signal from $M$=*32* noisy samples, $\delta_\varepsilon^2 = 0.05$. True (red) and estimated (gray) kernel positions and expert values. (c): Result of Multi-model fusion using "parameter averaging" with $H$=*10* models. Enlarge figure to see individual *(dots)* as well as fused parameters *(blue crosses)*. (Bottom): Gating functions.

## IV. MULTI-MODEL INFERENCE

We assume that images are sufficiently represented by a SMoE model with $L$ kernels and are corrupted with an additive noise signal $\varepsilon(\underline{x})$ from a covariance-stationary, zero-mean noise process $\{\mathcal{E}(\underline{x})\}$, $y_r(\underline{x}) = y(\underline{x}) + \varepsilon(\underline{x})$, $\mu_\varepsilon(\underline{x}) = 0$ and $\delta_\varepsilon^2(\underline{x}) = \delta_\varepsilon^2$. In general, also correlated noise can be assumed. With the SMoE mixture model we can account for varying noise variances at different „levels" of the signal y (i.e., suitable for Speckle Noise) $y_r(\underline{x}) = \sum_{j=1}^{L}\{m_j(\underline{x}) + \varepsilon_j(\underline{x})\} \cdot w_j(\underline{x})$. We shall assume $\sum_{j=1}^{L} \varepsilon_j(\underline{x}) \cdot w_j(\underline{x}) = \varepsilon(\underline{x})$, $\mu_{\varepsilon_j} = \mu_\varepsilon = 0$ and $\delta_{\varepsilon_j}^2 = \delta_\varepsilon^2$ for *j=1 ...L.*

For multi-model inference it is beneficial to analyze the role of the estimated parameters $\widehat{m}_j$. It can be shown that $m_j$ is estimated in region $R_j$ using

$$\widehat{m}_j = \hat{\mu}_{\hat{Y}_j} = \sum_{m=1}^{M} \hat{y}(\underline{x}_m) \cdot \frac{w_j(\underline{x}_m)}{\sum_{k=1}^{M} w_l(\underline{x}_k)}$$

We assume processes $\{\mathcal{E}(\underline{x})\}$ and $\{Y(\underline{x})\}$ are statistically independent, thus

$$\widehat{m}_j = \mathrm{m}_j + \hat{\mu}_{\varepsilon_j} \quad \text{and} \quad \delta_{\widehat{m}_j}^2 = \frac{\delta_\varepsilon^2}{M_j}$$

with $M_j = \sum_{l=1}^{M} w_j(\underline{x}_l)$. Notice that $\hat{\mu}_{\varepsilon_j} = \sum_{m=1}^{M} \varepsilon(\underline{x}_m) \cdot \frac{w_j(\underline{x}_m)}{M_j}$ is the bias for estimating the parameter $\mathrm{m}_j$ and is itself a random variable with expectation $\mathrm{E}[\hat{\mu}_{\varepsilon_j}] = 0$. Thus, with a sufficient number of noise samples $M_j$ captured by a gating function $w_j$ we can expect an unbiased estimate $\widehat{m}_j \rightarrow m_j$. $\delta_{\widehat{m}_j}^2$ measures our uncertainty about the bias which vanishes with $M_j \rightarrow \infty$. We conclude: **The larger the number of noise samples $\mathbf{M_j}$ covered by a gating function $\mathbf{w_j}$, the less biased the estimate of $m_j$, the more accurate the inference of the model.** Since $max\{M_j\}$ is restricted to small block size $M$, we tackle the "small sample" problem by searching for $H$ noisy signal blocks with same or similar original signals $y(\underline{x})$, albeit noisy.

For image denoising the strategy to find and use distant blocks with identical or similar content, superimposed with stochastically independent noise patterns, is used most prominently in the state-of-the-art BM3D algorithm. In contrast to BM3D we will denoise the $H$ identified individual noisy blocks with dedicated SMoE models $f^h(\underline{x}, \underline{\hat{\theta}}^h)$, optimized on the data $D^h = \{y_r^h(\underline{x}_m), \underline{x}_m\}_{m=1}^{M}$ of these blocks. Each of the $H$ models $f^h(\underline{x}, \underline{\hat{\theta}}^h)$ allows an inference on the signal $y(\underline{x})$. The prime question is how to use the $H$ hypotheses about the outcome for denoising. Important strategies include weighting the model outcomes $y(\underline{x}) = \sum_{h=1}^{H} f^h(\underline{x}, \underline{\hat{\theta}}^h) \cdot \Delta^h$ or weighting the parameters $\underline{\hat{\theta}} = \sum_{h=1}^{H} \underline{\hat{\theta}}^h \cdot \Delta^h$ of the models, i.e. with Akaike weights $\Delta^h$ based on validity of each model [30]. In our approach we can devise possible multi-model fusion strategies based on above findings:

*Averaging Parameters* - $\delta_{\widehat{m}_j^q}^2 = \delta_{\widehat{m}_j^p}^2$: If it is possible to identify $H$ image blocks with identical or at least very similar content we can assume all gating functions $w_j^h$ are approximately identical and thus $\delta_{\widehat{m}_j^q}^2 = \delta_{\widehat{m}_j^p}^2$ for any models $p$ and $q$ applies. Averaging the $H$ individual model parameters $\widehat{\mathrm{m}}_j = \frac{1}{H}\sum_{h=1}^{H} \widehat{\mathrm{m}}_j^h$ harvests $H \cdot M_j$ noise samples, resulting in a more reliable estimate with reduced variance $\delta_{\widehat{m}_j}^2 = \frac{\delta_\varepsilon^2}{H \cdot M_j}$. The expected maximum multi-hypothesis-gain $G_{Mul}^H$ via single model inference accounts to $G_{Mul}^H = H$. It is expected, that also averaging the entire parameter sets $\underline{\hat{\theta}} = \frac{1}{H}\sum_{h=1}^{H} \underline{\theta}^h$ achieves significant advantages. See Fig. 2 for illustration how this strategy can improve reconstruction quality.

*Averaging Predictions of Non-Aligned Blocks* - $\delta_{\widehat{m}_j^q}^2 \neq \delta_{\widehat{m}_j^p}^2$: Averaging inferences from $H$ models $\hat{y}(\underline{x}_k) = \sum_{h=1}^{H} f^h(\underline{x}_k, \underline{\hat{\theta}}^h) \cdot \Delta^h$ that support prediction of a pixel location $\underline{x}_k$ is also possible. An unbalanced weighting $\Delta^h$ of the individual models may be advisable, because the individual $\widehat{\mathrm{m}}_j^h$ parameters are based on different reliability outcomes $\delta_{\widehat{m}_j^q}^2 \neq \delta_{\widehat{m}_j^p}^2$. Such weighting may be based on reliability measures such as $\delta_{\widehat{m}_j^h}^2$ or Wiener filter weights as with BM3D.

## V. RESULTS FOR SMOE MMI

We investigated a block-based SMoE denoising approach using MMI on *M=8x8* blocks - to be able to integrate and compare with BM3D with the same block size. Full SMoE model in Eq. (2) was used with *L=4* steering kernels. GD optimization was employed for $\underline{\theta}$ in each block using stochastic gradient descent with mean-square-error as the objective function. All BM3D results reported are "best PSNR results" achieved with varying levels of hyperparameter *SIGMA* and based on software in [31].

### A) MMI from Local Neighborhoods (S-SMoE)

Inference blocks are found by shifting an *8x8* block window over the image from left to right and top to bottom, with shifts of *s* pixels, and retrieving the *M=8x8* noisy pixels in each region. Each block is denoised using a single SMoE model. After all blocks are processed, each pixel is MMI denoised by averaging denoised pixels from all models available at same locations (S-SMoE).

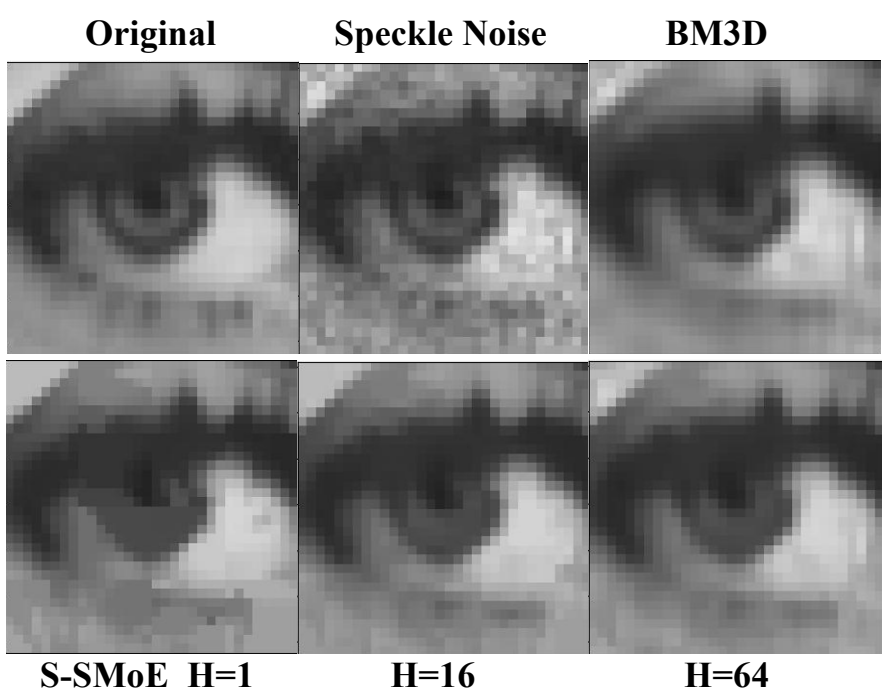

**Fig. 3.** Local Multi-Model Denoising: S-SMoE with 1, 16 and 64 models fused/pixel. Comparison with "non-local" BM3D.

Results with speckle noise ($\delta_\varepsilon^2 = 0.01$) in Fig. 3 show that the single SMoE model regression with non-overlapping blocks (*H=1*, thus *s=8*) is not sufficient. Fusing outcomes of many models significantly improves PSNR, SSIM as well as visual quality - as expected increasing with increasing *H* according to theory. PSNR and SSIM are provided in Tab. I. We achieve up to 0.7 dB PSNR gain compared to the best results of state-of-the art BM3D on this example, for both Gaussian and Speckle noise. BM3D is a non-local approach designed to harvests correlations from more distant blocks. For this image apparently few corresponding blocks for 3D groups are found which may explain the performance gain of S-SMoE. Also, BM3D seems to provide overly smoothed results, which may be due to the shrinkage of the 3D transform. On the contrary, thanks to the edge-accuracy of the SMoE regression model, the edges are remarkably preserved - even when fusing the outcomes of $H=64$ models/pixel ($s=1$) from neighboring blocks. Block artifacts are as well completely eliminated with MMI.

TABLE I

LOCAL NEIGHBORHOOD DENOISING RESULTS - PSNR AND SSIM

| | **H=1** | **H=16** | **H=64** | |
|---|---|---|---|---|
| **Speckle Noise** | | | | |
| S-SMoE | 31.0 dB /0.94 | **32.1 dB** /0.96 | **32.8 dB** /0.97 | $\delta_\varepsilon^2 = 0.01$ |
| BM3D | | | 32.0 dB/ 0.97 | $\delta_\varepsilon^2 = 0.01$ |
| S-SMoE | 28.4 dB /0.92 | **29.5 dB** /0.94 | **29.7 dB** /0.95 | $\delta_\varepsilon^2 = 0.02$ |
| BM3D | | | 29.5 dB /0.95 | $\delta_\varepsilon^2 = 0.02$ |
| **Gaussian Noise** | | | | |
| S-SMoE | 31.1 dB /0.94 | **32.8 dB** /0.96 | **32.9 dB** / 0.97 | $\delta_\varepsilon^2 = 0.01$ |
| BM3D | | | 32.1 dB / 0.96 | $\delta_\varepsilon^2 = 0.01$ |
| S-SMoE | 28.1 dB /0.91 | 29.3 dB /0.94 | **30.2 dB** /0.95 | $\delta_\varepsilon^2 = 0.02$ |
| BM3D | | | 30.1 dB / 0.95 | $\delta_\varepsilon^2 = 0.02$ |

TABLE II

NON-LOCAL DENOISING RESULTS - PSNR AND SSIM

| **Image** | **(1)** | **(2)** | **(3)** | **(4)** | **(5)** |
|---|---|---|---|---|---|
| **Speckle Noise, $\delta_\varepsilon^2 = 0.01$** | | | | | |
| S-SMoE *H=64* | **32.8 dB** 0.97 | **33.2 dB** 0.95 | **35.5 dB** 0.91 | **29.9 dB** 0.97 | **30.9 dB** 0.8 |
| BM-SMoE | **33.1 dB** 0.96 | 32.4 dB 0.94 | **35.6 dB** 0.93 | **30.2 dB** 0.96 | 30.6 dB 0.84 |
| BM3D | 32.0 dB 0.96 | 33.0 dB 0.95 | 35.1 dB 0.93 | 29.8 dB 0.96 | 30.8 dB 0.85 |

### B) Non-Local SMoE MMI (BM-SMoE)

BM3D very efficiently removes noise using collaborative filtering of blocks in 3D groups. Block-matching (BM) is used in BM3D to identify similar blocks in non-local neighborhoods. When significant edges are present in blocks the algorithm avoids overambitious noise reduction in 3D-transform domain – in order to avoid deformation of edges. The edge-aware SMoE model in contrast attempts to exactly reconstruct the details of such edges. We may expect the approach to perform especially well where BM3D may be underperforming.

For non-local BM-SMoE regression the MMI approach was integrated into the BM3D software [31] into the 1st step of BM3D algorithm. SMoE denoising was performed on each 8x8 block in a 3D group with max. 16 blocks/group. No 3D transform was employed. Images were reconstructed using BM3D procedure by averaging for each pixel all related SMoE denoised blocks from all 3D groups. The 2nd step of BM3D was not used. BM3D results were generated using identical BM as with BM-SMoE, in contrast using both 1st and 2nd step of the algorithm.

Fig. 4 and Tab. II show results on noisy test images. In general, S-SMoE produced comparable or superior results compared to BM3D with impressive edge sharpness close to originals. BM-SMoE results are less conclusive, but surprisingly for image (1) BM-SMoE produces similar high gains as with S-SMoE. More investigation into the subject is required to understand why particular gains materialize. Fig. 5 shows how the local S-SMoE approach with fusion of 64 models provides excellent results on "Lena" corrupted by heavy speckle noise $\delta_\varepsilon^2 = 0.02$.

## VI. CONCLUSION

Results show that the Multi-Model SMoE Inference approach can provide an interesting path for denoising of images. The ability of SMoEs to efficiently reconstruct varieties of edges from noise enables efficient inference with many models. Many inferior "biased" SMoE regressors from blocks with few samples combined produce superior, edge preserving, results. Fusion of even higher number of models from "local" S-SMoE as well as "global" BM-SMoE may provide an interesting path for further improvements. As our theoretical work suggests a simple "averaging" of model inferences may not provide optimal performance. After all, fusion gains in S-SMoE approach seem to saturate towards $H=64$ models in our experiments. As theory suggests, a weighting of the models according to the individual reliability appears promising, i.e. based on bias variances $\delta^2_{\hat{m}_j^q}$. Computational demands imposed by iterative optimization of the SMoE parameters are significant. Recent work shows that SMoE Autoencoders can be designed to estimate model parameters. Progress in this domain will allow ultra-fast MMI SMoE regression in near future with many models.

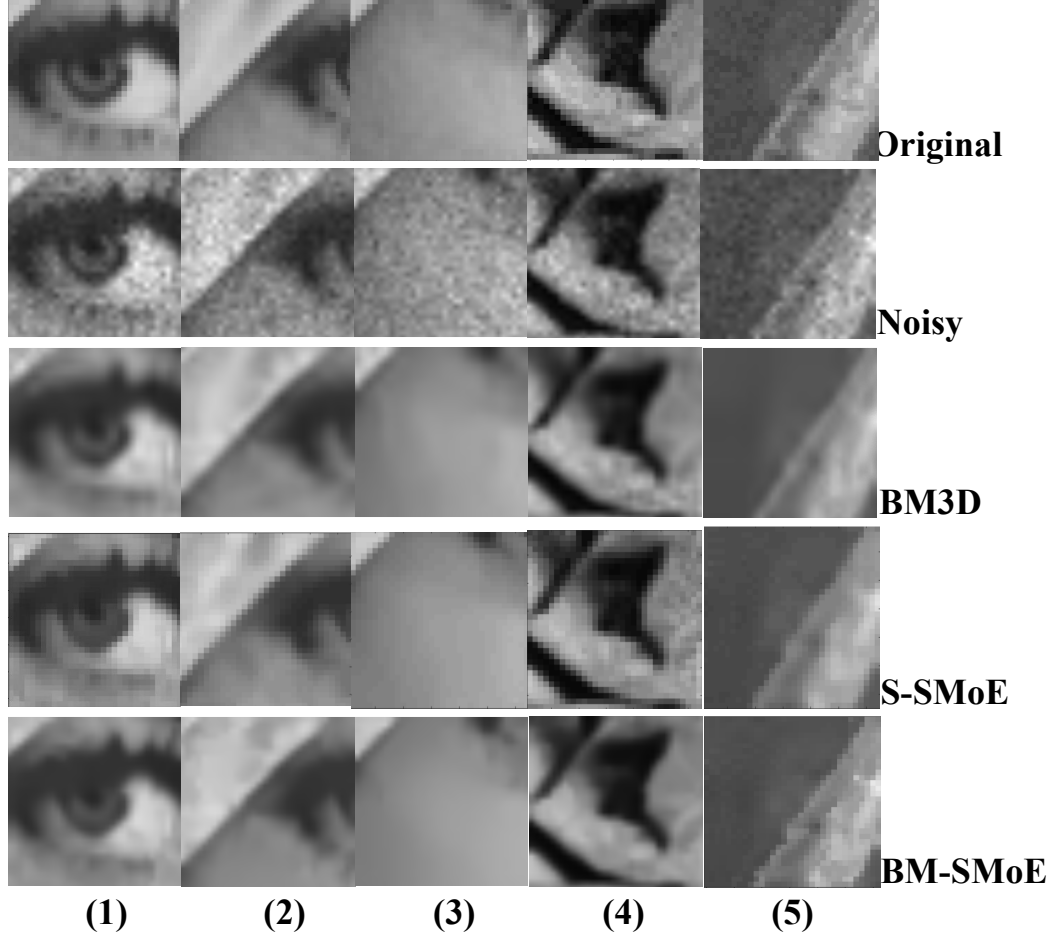

**Fig. 4.** BM-SMoE and BM3D on Speckle Noise, $\delta_\varepsilon^2 = 0.01$.

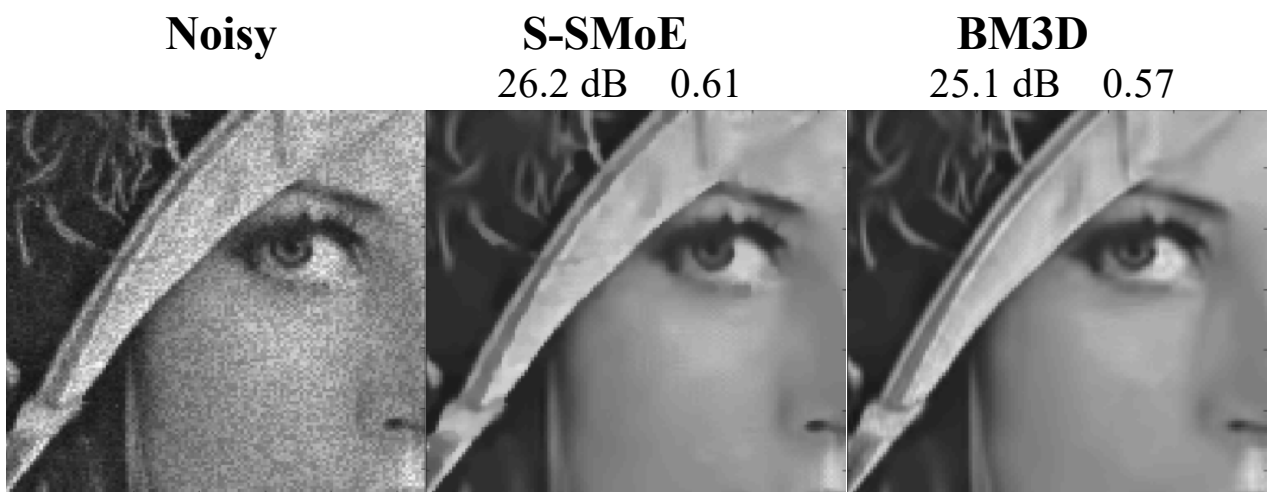

**Fig. 5.** Denoising of Lena with Speckle Noise, $\delta_\varepsilon^2 = 0.02$. S-SMoE with *H*=64 fused models. BM3D with "best PSNR" results.